\documentclass[aps,pra,twocolumn,superscriptaddress,nofootinbib,showpacs,amsmath,amssymb,floatfix]{revtex4}
\usepackage{graphicx}   
\usepackage{xspace}     
\usepackage{mathrsfs}
\newcommand{\notes}[1]{}
\newcommand{\xmath}[1]{\ensuremath{#1}\xspace}
\newcommand{\mrm}  [1]{\mathrm{#1}}
\newcommand{\mbf}  [1]{\mathbf{#1}}
\newcommand  {\Arg}{\mrm{Arg}}
\newcommand{\SSzero}{\xmath{\mrm{ {}^1S_0 }}}       
\newcommand{\TDone} {\xmath{\mrm{ {}^3D_1 }}}       
\newcommand{\TDTwo} {\xmath{\mrm{ {}^3D_2 }}}       
\newcommand{\TPone} {\xmath{\mrm{ {}^3P_1 }}}       
\newcommand{\TPzero}{\xmath{\mrm{ {}^3P_0 }}}       
\newcommand{\SPone} {\xmath{\mrm{ {}^1P_1 }}}       
\newcommand{\SSZero}{\xmath{\mrm{ 6s^2 }\:\SSzero }}
\newcommand{\TDOne} {\xmath{\mrm{ 5d6s }\:\TDone  }}
\newcommand{\TPZero}{\xmath{\mrm{ 6s6p }\:\TPzero }}
\newcommand{\SPOne} {\xmath{\mrm{ 6s6p }\:\SPone  }}
\newcommand{\SLJ}{\xmath{ {}^{2S+1}L_J }}
\newcommand{\energy}{\omega}
\newcommand{\SSZeroToTDOne}{\xmath{\SSZero~\rightarrow~\TDOne}}
\newcommand{\SSZeroToTDTwo}{\xmath{\SSZero~\rightarrow~\TDTwo}}

\newcommand{\AEONE} {\xmath{ A^{\mrm{(E1)   }} }}
\newcommand{\ASTARK}{\xmath{ A^{\mrm{(Stark)}} }}
\newcommand{\APV}   {\xmath{ A^{\mrm{(APV)  }} }}
\newcommand{\AMONE} {\xmath{ A^{\mrm{(M1)   }} }}
\newcommand{\HAtomic}{H_{\mathrm{Atomic}}}
\newcommand{\HZeeman}{H_{\mathrm{Zeeman}}}
\newcommand{\HStark}{H_{\mathrm{Stark}}}
\newcommand{\HAPV}{H_{\mathrm{APV}}}
\newcommand{\elight}{\xmath{\mathcal{E}}}

\newcommand{\klight}{\xmath{k}}
\newcommand{\efield}{\xmath{E}}
\newcommand{\bfield}{\xmath{B}}
\newcommand{\Elight}{\xmath{\boldsymbol{\elight}}}
\newcommand{\Klight}{\xmath{\mbf{\klight}}}
\newcommand{\Efield}{\xmath{\mbf{\efield}}}
\newcommand{\Bfield}{\xmath{\mbf{\bfield}}}

\newcommand{\Edc} {\xmath{\efield_{\mrm{dc}}}}
\newcommand{\Eac} {\xmath{\tilde{\efield}_{\mrm{0}}}}
\newcommand{\edc}{\xmath{e}}
\newcommand{\eac}{\xmath{\tilde{e}}}
\newcommand{\bdc}{\xmath{b'}}
\newcommand{\bac}{\xmath{\tilde{b}}}
\newcommand{\edipole}{\xmath{d}}
\newcommand{\mdipole}{\xmath{\mu}}
\newcommand{\MONE}{\mathcal{M}}
\newcommand{\Edipole}{\xmath{\mbf{\edipole}}}
\newcommand{\Mdipole}{\xmath{\boldsymbol{\mdipole}}}
\newcommand{\J}{\xmath{\mathbf{J}}}
\newcommand{\xhat}{\xmath{\hat{\mathbf{x}}}}
\newcommand{\yhat}{\xmath{\hat{\mathbf{y}}}}
\newcommand{\zhat}{\xmath{\hat{\mathbf{z}}}}
\newcommand{\nhat}{\xmath{\hat{\mathbf{n}}}}
\newcommand{\bra}[1]{\langle#1|}
\newcommand{\ket}[1]{|#1\rangle}
\newcommand{\clebsch}[6]{\xmath{\langle #1,#2;#3,#4|#5,#6\rangle}}

\newcommand{\kq}[3]{\xmath{#1^{#2}_{#3}}}
\newcommand{\reducedME}[4]{\xmath{(#1||#2^{#3}||#4)}}
\newcommand{\Rate}{\xmath{R}}
\newcommand{\ModRate}[2]{\xmath{\kq{\mathcal{\Rate}}{[#1]}{#2}}}
\newcommand{\ratio}{\xmath{\mathcal{K}}}
\newcommand{\ratioStark}{\xmath{r^{(\mrm{Stark})}}}
\newcommand{\ratioAPV}  {\xmath{r^{(\mrm{APV})}}}
\newcommand{\ratioMONE} {\xmath{r^{(\mrm{M1})}}}
\newcommand{\ratioEllip}{\xmath{r^{(\mrm{\phi})}}}
\newcommand{\vzmp}{\xmath{v_0}}
\newcommand{\kB}  {\xmath{k_B}}
\newcommand{\muB} {\xmath{\mu_{\mathrm{B}}}}

\newcommand\T{\rule{0pt}{4ex}}          
\newcommand\B{\rule[-3ex]{0pt}{0pt}}
\newcommand\smallB{\rule[-2ex]{0pt}{0pt}}
\begin{document}
%
%
\title{Parity violation in atomic ytterbium: experimental sensitivity and systematics}
\author{K. Tsigutkin}
\email{tsigutkin@berkeley.edu}
\author{D. Dounas-Frazer}
\author{A. Family}
\author{J. E. Stalnaker}
\altaffiliation[Present address: ]{Department of Physics and
Astronomy, Oberlin College, Oberlin, OH 44074}
\affiliation{Department of Physics, University of California at Berkeley,\\
Berkeley, CA 94720-7300}
\author{V. V. Yashchuk}
\affiliation{Advanced Light Source Division, Lawrence Berkeley
National Laboratory, Berkeley CA 94720}
\author{D. Budker}
\affiliation{Department of Physics, University of California at Berkeley,\\
Berkeley, CA 94720-7300}
\affiliation{Nuclear Science Division, Lawrence Berkeley National Laboratory, Berkeley, California 94720}
\date{\today}
%

\begin{abstract}
We present a detailed description of the observation of parity
violation in the \SSZeroToTDOne 408-nm forbidden transition of
ytterbium, a brief report of which appeared earlier. Linearly
polarized 408-nm light interacts with Yb atoms in crossed \Efield-
and \Bfield-fields. The probability of the 408-nm transition
contains a parity violating term, proportional to
$(\Elight\cdot\Bfield)[(\Efield\times\Elight)\cdot\Bfield]$,
arising from interference between the parity violating amplitude
and the Stark amplitude due to the E-field (\Elight is the
electric field of the light). The transition probability is
detected by measuring the population of the \TPzero state, to
which 65\% of the atoms excited to the \TDone state spontaneously
decay. The population of the \TPzero state is determined by
resonantly exciting the atoms with 649-nm light to the ${\rm 6s7s}
\:^3{\rm S}_1$ state and collecting the fluorescence resulting
from its decay. Systematic corrections due to E-field and B-field
imperfections are determined in auxiliary experiments. The
statistical uncertainty is dominated by parasitic frequency
excursions of the 408-nm excitation light due to imperfect
stabilization of the optical reference with respect to the atomic
resonance. The present uncertainties are 9\% statistical and 8\%
systematic.  Methods of improving the accuracy for the future
experiments are discussed.
\end{abstract}
%
\maketitle
%

\section{Introduction}
\label{Sect:Intro} In an earlier paper \cite{APV_PRL}, we reported
on observation of the atomic parity violation (PV) effect in the
\SSZeroToTDOne 408-nm forbidden transition of $^{174}$Yb. We
measured the PV induced transition matrix element to be $8.7\pm
1.4\times 10^{-10}$~ea$_0$, which confirms the theoretically
anticipated PV enhancement in Yb \cite{demille95} and constitutes
the largest atomic parity violation effect observed so far.
However, the measurement accuracy is not yet sufficient for the
observation of the isotopic and hyperfine differences in the PV
amplitude, the study of which is the main goal of the present
experiments.  Here we describe the impact of the apparatus
imperfections and systematic effects on the accuracy of the
measurements and discuss ways of improving it.

During the initial stage of the experiment, an effort was invested
into measuring various spectroscopic properties of the Yb system
of direct relevance to the PV measurement, including determination
of radiative lifetimes, measurement of the Stark-induced
amplitudes, hyperfine structure, isotope shifts, and dc-Stark
shifts of the \SSZeroToTDOne transition \cite{Bow96}. In addition,
the \SSZeroToTDTwo transition at 404 nm has been observed, and the
electric quadrupole transition amplitude and tensor transition
polarizability have been measured \cite{Bow99}. The forbidden
magnetic-dipole (M1) amplitude of the 408~nm transition was
measured to be $1.33\times 10^{-4}$~\muB using the
M1-(Stark-induced)E1 interference technique \cite{Sta2002}. The
ytterbium atomic system, where transition amplitudes and
interferences are well understood, has proven useful for gaining
insight into the Jones-dichroism effects that had been studied in
condensed-matter systems at extreme conditions and whose origin
had been a matter of debate (see Ref. \cite{Bud2003Jones} and
references therein).

An experimental and theoretical study of the dynamic (ac) Stark
effect on the \SSZeroToTDOne forbidden transition was also
undertaken \cite{Sta2006}. A model was developed to calculate
spectral line shapes resulting from resonant excitation of atoms
in an intense standing light wave in the presence of off-resonant
ac-Stark shifts. A bi-product of this work was an independent
determination (from the saturation behavior of the 408-nm
transition) of the Stark transition polarizability, which was
found to be in agreement with the earlier measurement
\cite{Bow99}.

The present Yb APV experiment involves a measurement using an
atomic beam. An alternative approach would involve working with a
heat-pipe-like vapor cell. Various aspects of such an experiment
were investigated, including measurements of collisional
perturbations of relevant Yb states \cite{Kim99}, nonlinear
optical processes in a dense Yb vapor with pulsed UV-laser
excitation \cite{DeB2001}, and an altogether different scheme for
measuring APV via optical rotation on a transition between excited
states \cite{Kim2001}.

The present paper addresses the issues of sensitivity and
systematics in the Yb APV experiment. In
Sections~\ref{Sect:Expt_Technique} and \ref{Sec:PV_Signature_Ideal
case} the experimental technique and its application in the
present experiment are discussed. In
Section~\ref{Sec:PV_signature_imperfections} a method of analyzing
the impact of various apparatus imperfections is described based
on theoretical modeling of signals recorded by the detection
system in the presence of imperfections. In
Section~\ref{Sec:Expt_System} a detailed description of the
experimental apparatus is given, along with a discussion of the
origins of the imperfections, which is followed by an account of
the measurements of the imperfections in
Section~\ref{Sec:Results_Analysis}. In
Sections~\ref{Sec:Error_budget} and \ref{Sec:Future} we discuss
measurements and analysis of the PV amplitude and systematic
effects, and ways of improving the accuracy of the PV measurements
to better than 1\% in order to measure the difference in the APV
effects between different isotopes and hyperfine components.

\section{Experimental technique for the APV measurement}
\label{Sect:Expt_Technique} 
\begin{figure}[!htb]
\resizebox{0.45\textwidth}{!}{%
  \includegraphics{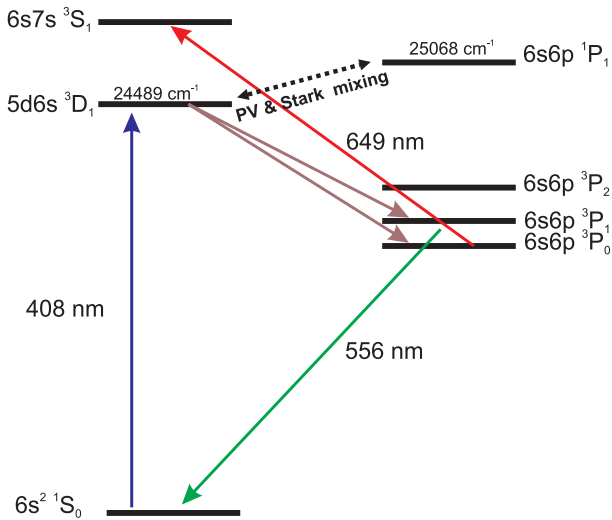}}
\caption{(color online) Low-lying energy eigenstates of Yb and
transitions relevant to the APV experiment.} \label{levels}
\end{figure}

As discussed in Ref.~\cite{APV_PRL},
the idea of the experiment is to excite the forbidden 408-nm
transition (Fig.~\ref{levels}) with resonant laser light in the
presence of a quasi-static electric field. The PV amplitude of
this transition arises due to PV mixing of the \TDOne and \SPOne
states. The purpose of the electric field is to provide a
reference transition amplitude, which is due to Stark mixing of
the same states interfering with the PV amplitude. In such an
interference method \cite{Bou75,Con79}, one is measuring the part
of the transition probability that is linear in both the reference
Stark-induced amplitude and the PV amplitude. In addition to
enhancing the PV dependent signal, the Stark-PV interference
technique provides for all-important reversals that separate the
PV effects from the systematics.

Even though the APV effect in Yb is relatively large, and the M1
transition is strongly suppressed, the M1 transition amplitude is
still three orders of magnitude larger than the PV amplitude. As a
result, the geometry of the experiment was designed to suppress
spurious M1-Stark interference. In addition, this effect is
minimized by the use of a power-build-up cavity to generate a
standing light wave. Since a standing wave has no net direction of
propagation any transition rate which is linear in the M1
amplitude, will cancel out (see below).

The advantages of the present experimental configuration can be
demonstrated by considering Yb atoms in the presence of static
electric, \Efield, and magnetic, \Bfield, fields interacting with
a standing monochromatic wave produced by two counter-propagating
coherent waves in an optical cavity. The electric field in the
standing wave, \Elight, is a sum of the fields of the two waves.
For resonant atoms, the transition rate from the ground state
$\SSzero$ to the excited state $\TDone$ is (see e.g.
\cite{RedBook}, Eq.~(3.127))
\begin{equation}\label{eq:R_M}
R_M = 
\frac{4}{\hbar^2\Gamma}|A_M|^2,
\end{equation}
where $\Gamma$ is the natural linewidth of the transition, $A_M$
is the transition amplitude, and $M=~0,\pm1$ is the magnetic
quantum number of the excited state. Here and in the rest of this
section it is assumed that the individual magnetic sublevels of
the $\TDone$ state are resolved. For convenience, we set $\hbar=1$
henceforth and measure the transition rate in units of $\Gamma$.

The transition amplitude $A_M$ is the sum of the electric- (E1)
and magnetic- (M1) dipole transition amplitudes:
\begin{equation}
A_M = \AEONE_M + \AMONE_M.
\end{equation}
The E1 amplitude has two contributions corresponding to the Stark-
and PV- mixing of the $\TDone$ and $\SPone$ states.  That is,
\begin{align}\label{StAmp_general}
\AEONE_M &= \ASTARK_M+\APV_M\nonumber\\
&=i\beta (-1)^{M}\left(\Efield\times\Elight\right)_{-M}+
i\zeta (-1)^{M}\elight_{-M},
\end{align}
where $\beta$ is the vector transition polarizability, $\zeta$ is
related to the reduced matrix element of the Hamiltonian
describing the weak interaction, and $\elight_{0,\pm1}$ are the
spherical components of the vector $\Elight$.  Although
Stark-induced transitions are generally characterized by scalar,
vector, and tensor polarizabilities \cite{Bou75,Bow99}, for the
case of a $J=0\rightarrow 1$ transition, only the vector
polarizability contributes.  Equation (\ref{StAmp_general}) is
derived in Appendix~\ref{Ap:amplitudes}.

Similarly, the M1 transition amplitude has two components: one for
each of the two counter-propagating laser beams.  Let
$\Elight_\mrm{+}=\elight_\mrm{+}\,\hat{\Elight}$ and
$\Elight_\mrm{-}=\elight_\mrm{-}\,\hat{\Elight}$ denote the
electric fields of the beams traveling in the $\Klight$ and
$-\Klight$ directions, respectively.  Then
$\Elight=\Elight_\mrm{+}+\Elight_\mrm{-}$ and the M1 amplitude is
given by
\begin{align}\label{M1Amp}
\AMONE_{M} &=
\MONE(-1)^{M}\left(\Klight\times\Elight_\mrm{+}\right)_{-M}+
\MONE(-1)^{M}
\left(-\Klight\times\Elight_\mrm{-}\right)_{-M}\nonumber\\
&= \MONE(-1)^{M}\left(\delta\Klight\times\Elight\right)_{-M},
\end{align}
where $\MONE$ is the reduced matrix element of the M1 transition
and $\Klight$ is a unit vector in the direction of the wavevector.
Here we have introduced $\delta\Klight= \delta\klight\,\Klight$
with $\delta\klight= (\elight_\mrm{+}-\elight_\mrm{-})/\elight$.
For a perfect standing wave, $\elight_\mrm{+}=\elight_\mrm{-}$ and
hence $\delta\klight=0$ and the M1 transition is completely
suppressed. In practice, $\elight_\mrm{-}=\elight_\mrm{+} -
\delta\elight$ due to the small but nonzero transmission of the
back mirror in the cavity.  Since $|\delta\elight|\ll\elight$,
$|\delta\klight|\approx |\delta\elight/\elight| \ll 1$. Thus,
although the M1 transition amplitude is not strictly zero, it is
highly suppressed.

Without loss of generality, the quantities $\beta$, $\zeta$, and
$\MONE$ are assumed to be real. In general,  the rate $R_M$ given
by Eq.~(\ref{eq:R_M}) includes terms proportional to $\beta\MONE$
(Stark-M1 interference) and $\beta\zeta$ (Stark-PV interference).

\begin{figure}[htb]
\resizebox{0.47\textwidth}{!}{%
  \includegraphics{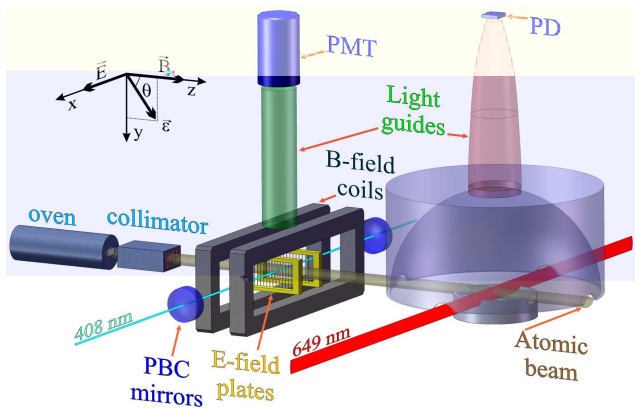}
} \caption{(color online) Orientation of fields for PV-Stark
interference experiment and schematic of the present APV
apparatus. Not shown is the vacuum chamber containing all the
depicted elements, except the photomultiplier (PMT) and the
photodiode (PD). PBC--power buildup cavity. Light is applied
collinearly with \textbf{x}.} \label{apparatus}
\end{figure}

A careful choice of field geometry allows for additional
suppression of undesirable Stark-M1 interference.  From
Eq.~(\ref{StAmp_general}), it is evident that the Stark-PV
interference is proportional to the rotational invariant
\begin{equation}\label{Eq:Invariant}
(\underbrace{\Elight}_{\mrm{PV}}\cdot\Bfield)
[\underbrace{(\Efield\times\Elight)}_{\mrm{Stark}}\cdot\Bfield],
\end{equation}
which is P-odd and T-even. In the present experimental apparatus
the electric field, \Efield, is applied orthogonally to the
magnetic field, \Bfield, and collinearly with the axis of the
linearly-polarized standing light wave, as shown in
Fig.~\ref{apparatus}.

This geometry is such that the M1 and Stark-induced amplitudes are
out of phase. Thus, they do not interfere and therefore do not
produce spurious PV-mimicking effects (see
Section~\ref{Sec:PV_signature_imperfections}).

\section{PV signature: Ideal case}
\label{Sec:PV_Signature_Ideal case} In the ideal case where we
neglect the apparatus imperfections, the static magnetic and
electric fields are $\Bfield = B\,\zhat$ and $\Efield = E\,\xhat$,
respectively, and the light standing wave has an electric field
\begin{equation}\label{E_light}
\Elight=\elight(\sin\theta\,\yhat+\cos\theta\,\zhat).
\end{equation}
With this field orientation (see Fig. \ref{apparatus}), Eqs.
(\ref{eq:R_M}) through (\ref{M1Amp}) yield
\begin{align}
\Rate_0 & =4\elight^2 \left(\beta^2\efield^2\sin^2\theta+
2\zeta\,\beta\efield\sin\theta\cos\theta\right),\label{even_rates}\\
\Rate_{\pm 1}&=2\elight^2 \left(\beta^2\efield^2\cos^2\theta-
2\zeta\,\beta\efield\sin\theta\cos\theta\right),\label{even_rates2}
\end{align}
where terms of order $\zeta^2$ and higher are neglected, and
$\delta\klight=0$ is assumed.

In order to isolate the Stark-PV interference term from the
dominant Stark-induced transition rate, we modulate the electric
field: $\efield = \Edc+\Eac\cos(\omega t)$, where \Eac is the
modulation amplitude, $\omega$ is the modulation frequency, and
\Edc provides a DC bias.  Then Eqs. (\ref{even_rates}) and
(\ref{even_rates2}) become
\begin{equation}
\Rate_M = \ModRate{0}{M} + \ModRate{1}{M}\cos(\omega t) +
\ModRate{2}{M}\cos(2\omega t),
\end{equation}
where $\ModRate{n}{M}$ is the amplitude of the $n$th harmonic of
the transition rate $\Rate_M$.  The dominant Stark-induced
contribution, which oscillates at twice the modulation frequency,
is
\begin{align}
\ModRate{2}{0} &= 2\beta^2
\Eac^2\elight^2\sin^2\theta,\\
\ModRate{2}{\pm1} &= \beta^2 \Eac^2\elight^2\cos^2\theta.
\end{align}
On the other hand, the amplitude $\ModRate{1}{M}$ contains the
Stark-PV interference term:
\begin{align}
\ModRate{1}{0} &= 8\elight^2\left( \beta^2\Eac\Edc\sin^2\theta+
\zeta\beta\Eac\sin\theta\cos\theta\right),\\
\ModRate{1}{\pm1} &= 4\elight^2\left( \beta^2\Eac\Edc\cos^2\theta-
\zeta\beta\Eac\sin\theta\cos\theta\right).
\end{align}
The term \ModRate{0}{M} is a constant ``background":
\begin{align}
\ModRate{0}{0} &= 4\elight^2\left(
\beta^2(\Eac^2+\Edc^2)\sin^2\theta+
4\zeta\beta\Edc\sin\theta\cos\theta\right),\nonumber\\
\ModRate{0}{\pm1} &= 2\elight^2\left(
\beta^2(\Eac^2+\Edc^2)\cos^2\theta-
4\zeta\beta\Edc\sin\theta\cos\theta\right).\nonumber
\end{align}

\begin{figure}[!htb]
\resizebox{0.47\textwidth}{!}{%
  \includegraphics{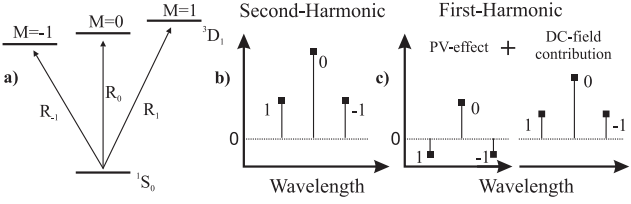}
} \caption{Discrimination of the PV effect by E-field modulation
under static magnetic field. The Zeeman pattern is shown for the
polarization angle $\theta=\pi/4$. } \label{lshape}
\end{figure}

For an arbitrary polarization angle $\theta$, all three Zeeman
components of the transition, as shown in Fig. \ref{lshape}a, are
present while scanning over the spectral profile of the
transition. The first-harmonic signal due to Stark-PV interference
has a characteristic signature: the sign of the oscillating terms
for the two extreme components of the transition is opposite to
that of the central component. The second-harmonic signal provides
a reference for the lineshape since it is free from interference
effects linear in \Efield (Fig. \ref{lshape}b). With a non-zero DC
component present in the applied electric field, a signature
identical to that in the second harmonic will also appear in the
first harmonic, Fig. \ref{lshape}c. The latter can be used to
increase the first-harmonic signal above the noise, which makes
the profile analysis more reliable.

To obtain the PV term from the measured first- and second-harmonic
transition rates, we first normalize the first-harmonic signals
\ModRate{1}{M} by their second-harmonic counterparts
\ModRate{2}{M} and combine the results in the following way:
\begin{equation}  \label{ratios}
\ratio= \frac{\ModRate{1}{-1}}{\ModRate{2}{-1}}+
\frac{\ModRate{1}{+1}}{\ModRate{2}{+1}}-
2\frac{\ModRate{1}{0}}{\ModRate{2}{0}} =
\mp\frac{16\zeta}{\beta\Eac}.
\end{equation}
Here we take $\theta=\pm\pi/4$, which are optimal polarization
angles for the PV measurements (see next section). This method has
the advantage that \ratio is independent of \Edc, so that the
E-field bias may be chosen based on technical requirements of the
experimental apparatus.

\section{PV signature: Impact of apparatus imperfections}
\label{Sec:PV_signature_imperfections} While the current Yb-APV
apparatus has been designed to minimize systematic effects, the PV
mimicking systematics may be a result of a combination of multiple
apparatus imperfections. In order to understand the importance of
these effects, the electric and magnetic field misalignments and
stray fields were included in a theoretical model of the
transition rates as well as the excitation light's deviations from
linear polarization. In addition, we relax the assumption that
$\delta\klight=0$ and include the effects of the residual M1
transition.

The quantization axis is defined along $\zhat$, and following the
ideal case model, the axis of the standing light wave is collinear
with $\xhat$. We added a small ellipticity to the light
polarization by taking
\begin{equation}  \label{ellip}
\Elight=\elight(\yhat\sin \theta+\zhat\,e^{i\phi}\cos \theta),
\end{equation}
where $\phi$ is a small phase. For $|\phi| \ll 1$, the ellipticity
of the light is $2\phi\sin(2\theta)$. The electric field
imperfections are included by taking
\begin{eqnarray}
\Efield = \tilde{\textbf{E}} + \Efield',\nonumber
\end{eqnarray}
where
\begin{eqnarray}
\tilde{\textbf{E}} &=& (\Eac\xhat+\eac_y\yhat+\eac_z\zhat)\cos(\omega t)\nonumber\\
\Efield' &=& \Edc\xhat+\edc_y\yhat+\edc_z\zhat,\nonumber
\end{eqnarray}
are the AC and DC components of the electric field. It is assumed
that the y- and z-components of the AC field are in phase with the
leading oscillating E-field. The impact of the out-of-phase AC
components was analyzed within a complete model of the systematics
and found to be negligible. The AC components are due to
misalignments of the applied E-field with respect to the light
wave axis as well as to the quantization axis $\zhat$. The DC
components arise due to a misalignment of the DC-bias field and
also due to stray electric fields in the interaction region.

The magnetic-field imperfections are defined within the same frame
of reference by taking analogously
\begin{eqnarray}
\Bfield = \tilde{\textbf{B}} + \Bfield',\nonumber
\end{eqnarray}
where
\begin{eqnarray}
\tilde{\textbf{B}} &=& \bac_x\xhat+\bac_y\yhat+\bfield\zhat\nonumber\\
\Bfield' &=& \bdc_x\xhat+\bdc_y\yhat+\bdc_z\zhat,\nonumber
\end{eqnarray}
where $\tilde{\textbf{B}}$ and $\Bfield'$ are reversing and stray
non-reversing magnetic fields, respectively.

Equations (\ref{eq:R_M}) through (\ref{M1Amp}) apply  when the
quantization axis is along the magnetic field, thus a rotation
$\mathbb{D}$ is applied to each of the vectors $\Efield$,
$\Bfield$, $\Elight$, and $\Klight$ such that
$\mathbb{D}\Bfield\propto\zhat$.  That is, we take
\begin{equation}
\Bfield\rightarrow \mathbb{D}\Bfield,\;\; \Efield\rightarrow
\mathbb{D}\Efield,\;\; \Elight\rightarrow
\mathbb{D}\Elight,\;\;\mrm{and}\;\; \Klight\rightarrow
\mathbb{D}\Klight,
\end{equation}
where
\begin{equation}
\mathbb{D} =
\mathscr{D}(-\alpha_y,\yhat)\mathscr{D}(\alpha_x,\xhat).
\end{equation}
Here the matrix $\mathscr{D}(\alpha,\nhat)$ represents a rotation
about an axis $\nhat$ through angle $\alpha$.  The angles
$\alpha_x$ and $\alpha_y$ are given by
\begin{equation}
\alpha_{x,y}  = (B-\bdc_z)(\bdc_{y,x}+\bac_{y,x})/B^2.
\end{equation}
Thus, the electric field \Efield and the polarization vector
\Elight acquire additional components after the rotation (besides,
for example, $\edc_y$ and $\eac_y$).

Due to the imperfections, the normalized-rate modulation
amplitudes now include additional terms besides the Stark- and the
PV effects:
\begin{equation}
\frac{\ModRate{1}{M}}{\ModRate{2}{M}} \equiv r_M = \ratioStark_M +
\ratioAPV_M + \ratioMONE_M + \ratioEllip_M,
\end{equation}
where $\ratioStark_M$ is the Stark contribution due to the DC-bias
and the field imperfections, $\ratioAPV_M$ is the
PV-Stark-interference term, $\ratioMONE_M$ is the
M1-Stark-interference contribution, and $\ratioEllip_M$ is a
contribution due to the distorted linear polarization of the light
(which is a Stark contribution, but we explicitly separate the
contribution linear in $\phi$). Expressions for the lowest-order
terms are summarized in the Table~\ref{imp_terms}.

\begin{table}[!htb]
\caption{Lowest-order terms contributing to the normalized
transition-rate modulation amplitudes $r_M$.} \label{imp_terms}
\begin{ruledtabular}
\begin{tabular}{l|c|c|c}
\smallB &
$\ratioAPV_M$ & $\ratioMONE_M$ & $\ratioEllip_M$ \\ \hline
$M=0$ \T\B &
$\displaystyle{+\frac{4\,\zeta\cot\theta}{\beta\Eac}}$ &
0 & 0 \\ \hline
$M=-1$ \T\B &
$\displaystyle{-\frac{4\,\zeta\tan\theta}{\beta\Eac}}$ &
$\displaystyle{+\frac{4\,\delta\klight\,
    \MONE(\tilde{e}_y-\tilde{e}_z\tan\theta)}
    {\beta\Eac^2}}$ &
$\displaystyle{+\frac{4\,e_z\phi\tan\theta}{\Eac}}$ \\ \hline
$M=+1$ \T\B &
$\displaystyle{-\frac{4\,\zeta\tan\theta}{\beta\Eac}}$ &
$\displaystyle{-\frac{4\,\delta\klight\,
    \MONE(\tilde{e}_y-\tilde{e}_z\tan\theta)}
    {\beta\Eac^2}}$ &
$\displaystyle{-\frac{4\,e_z\phi\tan\theta}{\Eac}}$ \\ \hline
\end{tabular}
\end{ruledtabular}
\end{table}

The normalized-amplitude combination (\ref{ratios}) has been
chosen to determine the PV asymmetry. Since the M1 and ellipticity
terms have opposite signs for $M=\pm1$, their contributions to
$\ratio$ cancel, while the contributions from $\ratioAPV_M$ add.

The Stark-contribution, $\ratioStark_M$, has several terms that
are produced due to different imperfections and impacts all three
Zeeman components, $M=0,\pm 1$. In order to determine which terms
could potentially mimic the PV asymmetry in $\mathcal{K}$, we
discriminate the PV contribution to $\mathcal{K}$ with respect to
the B-field reversal and flip of the polarization angle, $\theta$.
Switching to a different Zeeman component of the transition is
also a reversal, which is incorporated in the expression for the
asymmetry, $\mathcal{K}$. Analysis of the noise affecting the
accuracy of PV-asymmetry measurements demonstrate that the highest
signal-to-noise ratio is achieved when $\theta=\pm\pi/4$, and
therefore, the polarization flip is a change of the polarization
angle by $\pi/2$. Thus, the normalized-amplitude combination
(\ref{ratios}) must be determined for four different combinations
of the B-field directions and light-polarization angles:
$\mathcal{K}(+B,+\pi/4)$, $\mathcal{K}(-B,+\pi/4)$,
$\mathcal{K}(+B,-\pi/4)$, and $\mathcal{K}(-B,-\pi/4)$, so that
terms having different symmetries with respect to the reversals
can be isolated:
\begin{equation}
\label{Eq:reversals} \left[\begin{array}{c}
\ratio_1 \\ \ratio_2 \\
\ratio_3 \\ \ratio_4
\end{array}\right] =\frac{1}{4}
\left[\begin{array}{rrrr}
-1 & -1 & +1 & +1 \\
-1 & +1 & +1 & -1 \\
+1 & -1 & +1 & -1 \\
+1 & +1 & +1 & +1
\end{array}\right]\cdot
\left[\begin{array}{c}
\ratio(+B,+\theta) \\ \ratio(-B,+\theta) \\
\ratio(+B,-\theta) \\ \ratio(-B,-\theta)
\end{array}\right].
\end{equation}
The result of this procedure is summarized in
Table~\ref{reversals}.
\begin{table}[!htb]
\caption{List of the lowest-order terms contributing to the
asymmetry $\ratio$ for $|\theta|=\pi/4$ sorted with respect to
their response to the reversals. $\ratio_4$ corresponding to a
rather long list of terms that are invariant with respect to all
reversals, is not shown in the table.}
\label{reversals}%
\begin{ruledtabular}
\begin{tabular}{c|c|c}
$\ratio_1$ & $\ratio_2$ \smallB & $\ratio_3$
\\ \hline
$\displaystyle{
\frac{8(\eac_y\edc_z+\eac_z\edc_y)}{\Eac^2}+\frac{16 \bac_x
e_y}{B\Eac}+\frac{16\zeta}{\beta\Eac} }$  \T\B &
$\displaystyle{\frac{16\bdc_x\edc_y}{\bfield\Eac}}$ &
$\displaystyle{\frac{16\bdc_x\edc_z}{\bfield\Eac}}$
\end{tabular}
\end{ruledtabular}
\end{table}

The PV asymmetry contributing to $\ratio_1$ is B-field even,
$\theta$-flip odd. It competes with the second-order terms that
are a combination of the E-field and B-field alignment
imperfections and stray fields. Using the theoretical value of
$\zeta\simeq 10^{-9}~\text{ea}_0$ \cite{Por95,Das97} combined with
the measured $|\beta |= 2.24_{-0.12}^{+0.07}\times
10^{-8}~\text{ea}_0$/(V/cm) \cite{Bow99,Sta2006}, the expected PV
asymmetry, $16\zeta/\beta\tilde{E}$, is $\sim 4\cdot 10^{-4}$, for
$\theta=\pi/4$ and $\Eac=2$~kV/cm. For a typical value of
misalignments and ``parasitic" fields, $e_{y,z}/\Eac$ and
$\bac_{x}/\bfield$ (on the order of $10^{-3}$ in the present
apparatus), the contribution of the ``parasitic" terms may be up
to a few percent of the total value of $\ratio_1$. Ways of
measuring the contribution of these imperfections are discussed in
the following sections.

\section{Experimental apparatus}
\label{Sec:Expt_System} The forbidden 408-nm transition is excited
by resonant laser light coupled into the power-buildup cavity in
the presence of the magnetic and electric fields. The transition
rates are detected by measuring the population of the \TPzero
state, where 65\% of the atoms excited to the \TDone state decay
spontaneously (Fig. \ref{levels}). This is done by resonantly
exciting the atoms with 649-nm light to the ${\rm 6s7s} \:^3{\rm
S}_1$ state downstream from the main interaction region, and by
collecting the fluorescence resulting from the decay of this state
to the \TPone and $^3{\rm P}_2\:$ states and subsequently, from
the decay of the \TPone state to the ground state \SSzero (556~nm
transition). As long as the 408-nm transition is not saturated,
the fluorescence intensity measured in the probe region is
proportional to the rate of that transition.

A schematic of the Yb-APV apparatus is shown in Fig.
\ref{apparatus}. A beam of Yb atoms is produced (inside of a
vacuum chamber with a residual pressure of $\approx 3 \times
10^{-6}~{\rm Torr}$) with an effusive source: a stainless-steel
oven loaded with Yb metal, operating at $500-600^{\circ}{\rm C}$.
The oven is outfitted with a multi-slit nozzle, and there is an
external vane collimator reducing the spread of the atomic beam in
the horizontal direction. The resulting Doppler width of the
408-nm transition is $\approx 12 \, {\rm MHz}$ \cite{Sta2006}.

\begin{figure}[!htb]
\resizebox{0.47\textwidth}{!}{%
  \includegraphics{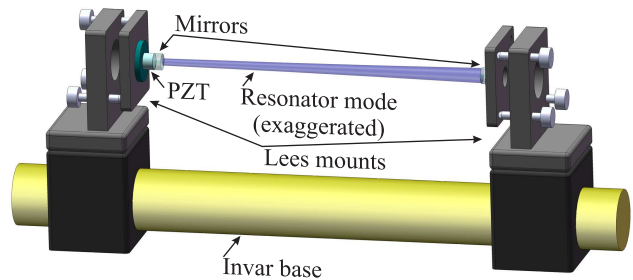}
} \caption{(color online) Schematic of the power buildup cavity.}
\label{pbc}
\end{figure}
Downstream from the collimator, the atoms enter the main
interaction region where the Stark- and PV-induced transitions
take place. Up to 80~mW of light at the transition wavelength of
$408.345 \, {\rm nm}$  in vacuum is produced by frequency doubling
the output of a Ti:Sapphire laser (Coherent $899$) using the
Wavetrain$^{\mbox{cw}}$ ring-resonator doubler. After shaping and
linearly polarizing the laser beam, $\approx 10$~mW of the 408-nm
radiation is coupled into a power buildup cavity (PBC) inside the
vacuum chamber.

\begin{figure*}[htb]
\resizebox{0.7\textwidth}{!}{%
  \includegraphics[bb=100 320 550 700]{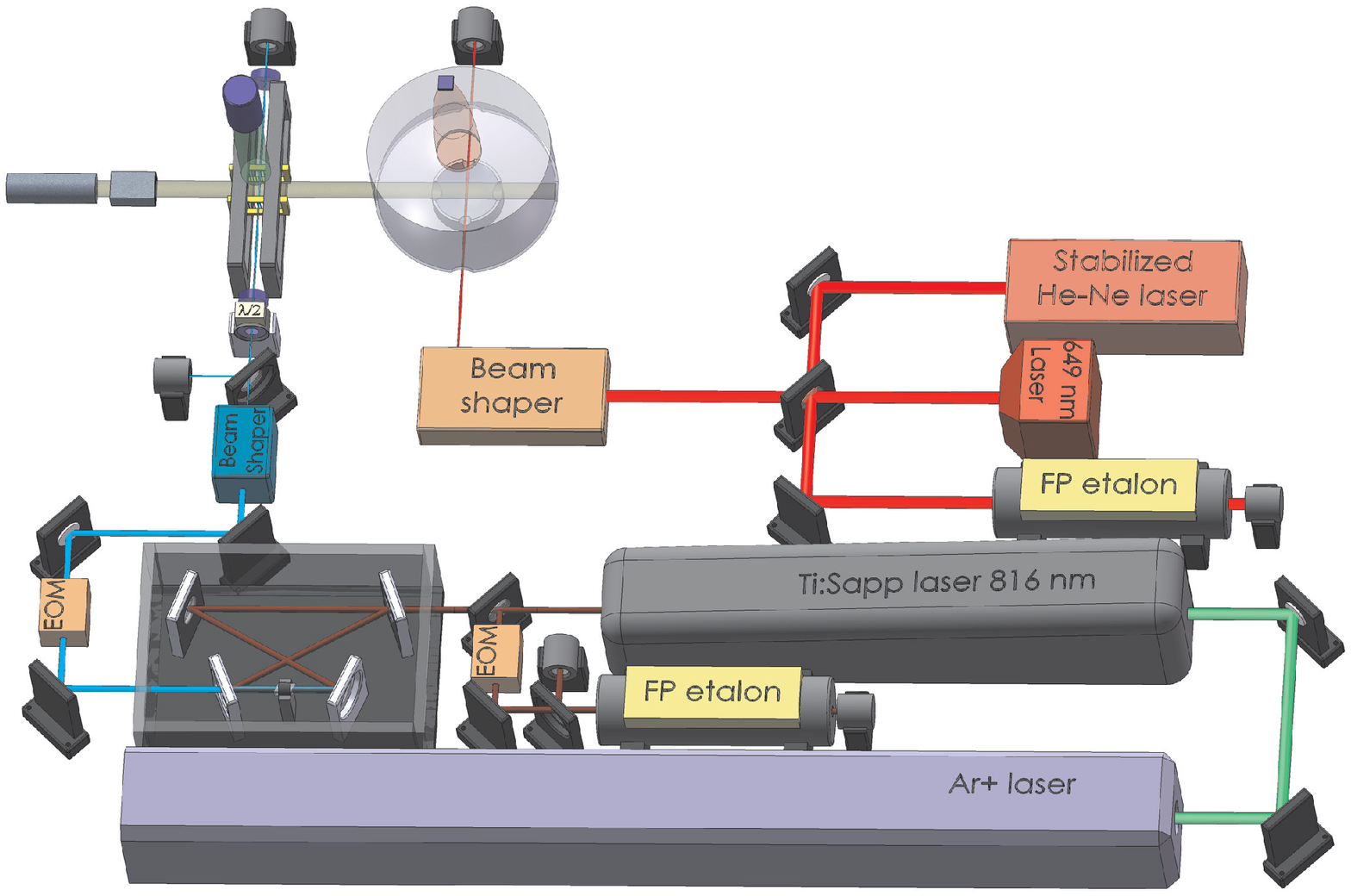}
} \caption{(color online) Schematic of the optical setup. Light at
408-nm is produced by frequency doubling the output of a
Ti:Sapphire laser (Coherent $899$) using the
Wavetrain$^{\mbox{cw}}$ ring resonator doubler. The laser is
locked to the PBC using the FM-sideband technique. The PBC is
locked to a confocal Fabry-P\'{e}rot \'{e}talon. This scannable
\'{e}talon provides the master frequency. The 649-nm excitation
light is derived from a single-frequency diode laser (New Focus
Vortex). The diode laser is locked to a frequency-stabilized He-Ne
laser using another scanning Fabry-P\'{e}rot \'{e}talon.}
\label{optics}
\end{figure*}
The cavity was designed to operate as an asymmetric cavity with
flat input mirror and curved back mirror with a 25-cm radius of
curvature and 22-cm separation between the mirrors. The atomic
beam intersects the cavity mode in the middle of the cavity, where
the $1/e^2$ radius of the mode in intensity is 172~$\mu\mbox{m}$.
The asymmetric configuration has the advantage of larger mode
radius at the interaction position compared to a symmetric cavity.
A larger mode allows us to reduce the ac-Stark shifts,
consequently reducing the width of the 408~nm transition.
Alternatively, the cavity can be modified to operate in the
symmetric confocal condition where multiple transverse modes can
be excited, thereby increasing the effective ``mode" size.
However, we were unable to obtain high power and stable lock for
the confocal configuration.

The cavity mirrors were purchased from Research Electro Optics,
Inc. For the flat input mirror the transmission is 350~ppm with
the absorption and scattering losses of 150~ppm total at 408~nm.
The curved back mirror is designed to have a lower transmission of
50~ppm in order to additionally suppress the net light wave vector
and, therefore, the M1 transition amplitude. The absorption and
scattering losses in the curved mirror are 120~ppm. The finesse
and the circulating power of the PBC are up to $\mathcal{F}=9000$
and $P=8$~W. These parameters were routinely monitored during the
PV measurements. Details of the characterization of the PBC are
addressed in Appendix \ref{Ap:PBC_mirrors}.

We found that the use of the 408-nm-PBC in vacuum is accompanied
by substantial degradation of the mirrors. Typically after 6 hours
of operation, the finesse drops by a factor of two. This is a
limiting factor for the duration of the measurement run. The
degradation of the finesse is due to the increased absorption and
scattering losses. This effect is reversible: the mirror
parameters can be restored by operating the PBC for several
minutes in air, which makes performing a number of runs possible
without replacing the mirrors. However, it takes several hours
with the present apparatus to reach the desired vacuum after
exposing the PBC to air. Presently, this effect is under
investigation aiming for longer-duration experiments and shorter
breaks in between.

\begin{figure}[htb]
\resizebox{0.47\textwidth}{!}{%
  \includegraphics{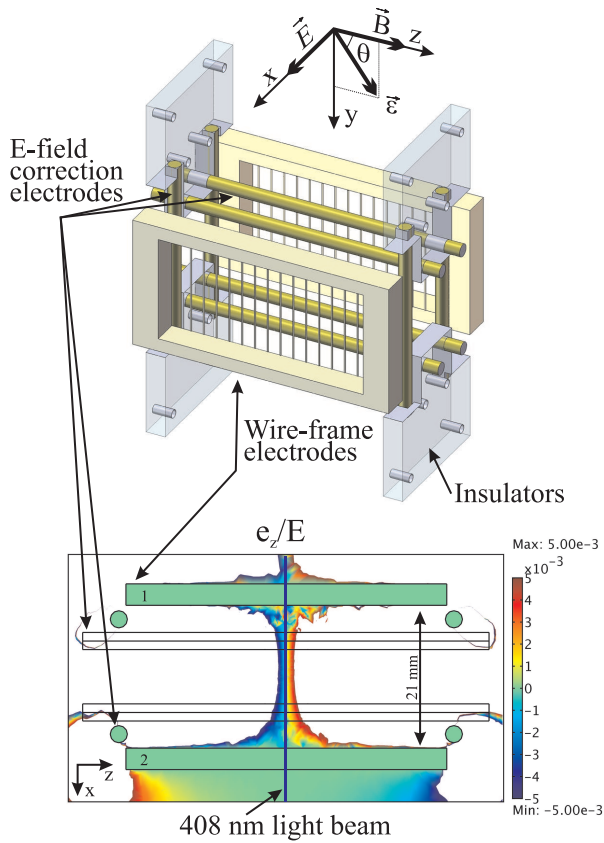}
} \caption{(color online) Schematic of the E-field electrodes
assembly, and a result of the E-field modeling showing an X-Z
slice of the amplitude of the E-field z-component, $e_z$, in a
midplane (Y=0) of the assembly normalized by the total E-field
amplitude, E. The voltage is applied to electrode 1, and electrode
2 and the correction electrodes are grounded.} \label{electrodes}
\end{figure}
A schematic of the PBC setup is presented in Fig. \ref{pbc}. The
mirrors are mounted on precision optical mounts (Lees mounts) with
micrometer adjustments for the horizontal and vertical angles and
the pivot point of the mirror face. The mirror mounts are attached
to an Invar rod supported by adjustable table resting on lead
blocks. The input mirror is mounted on a piezo-ceramic transducer
allowing cavity scanning.

The laser is locked to the PBC using the FM-sideband technique
\cite{Drev83}. In order to remove frequency excursions of the PBC
in the acoustic frequency range, the cavity is locked to a more
stable confocal Fabry-P\'{e}rot \'{e}talon, once again using the
FM-sideband technique.  This stable scannable cavity provides the
master frequency, with the power-build-up cavity serving as the
secondary master for the laser. A schematic of the optical system
is presented in Fig. \ref{optics}.

The magnetic field is generated by a pair of rectangular coils
designed to produce a magnetic field up to $100\ $G with a 1\%
non-uniformity over the volume with the dimensions of $1\times
1\times 1~\text{cm}^3$ in the interaction region. Additional coils
placed outside of the vacuum chamber compensate for the external
magnetic fields down to $10~\text{mG}$ at the interaction region.
The residual B-field of this magnitude does not have an impact on
the PV measurements since its contribution is discriminated using
the field reversals.

The electric field is generated with two wire-frame electrodes
separated by 2.1~cm (see Fig.~\ref{electrodes}). The copper
electrode frames support arrays of 0.2-mm diameter gold-plated
wires. This design allows us to reduce the stray charges
accumulated on the electrodes by minimizing the surface area
facing the atomic beam, thereby minimizing stray electric fields.
AC voltage of up to 10~kV at a frequency of 76.2~Hz is supplied to
the electrodes via a high-voltage amplifier. An additional DC bias
voltage of up to 100~V can be added.

The result of the electric field non-uniformity calculations is
shown in Fig.~\ref{electrodes}. These calculations demonstrate
that errors in the centering of the light beam with respect to the
E-field plates may induce substantial parasitic components as
large as, for example, $|e_z|\sim 5\times 10^{-3}\Eac$, producing
parasitic effects comparable to the PV asymmetry. In order to
measure and/or compensate the impact of the parasitic fields,
additional electrodes designed to simulate stray E-field
components were added to the interaction region. By applying
high-voltage to these electrodes (``correction electrodes" in
Fig.~\ref{electrodes}), the parasitic-field components may be
exaggerated and accurately measured as described in the following
sections.

Light at 556~nm emitted from the interaction region is collected
with a light guide and detected with a photomultiplier tube. This
signal is used for initial selection of the atomic resonance and
for monitoring purposes. Fluorescent light from the probe region
is collected onto a light guide using two optically polished
curved aluminum reflectors and registered with a cooled
photodetector~(PD). The PD consists of a large-area ($1\times
1~\text{cm}^2$) Hamamatsu photodiode connected to a 1-G$\Omega$
transimpedance pre-amplifier, both contained in a cooled housing
(temperatures down to $-15^{\circ}{\rm C}$). The pre-amp's
bandwidth is 1~kHz and the output noise is $\sim 1$~mV (rms). The
649-nm excitation light is derived from a single-frequency diode
laser (New Focus Vortex) producing $\approx 1.2\ $mW of cw output,
high enough to saturate the {\mbox{${\rm 6s6p \:} ^3{\rm
P}_0~\rightarrow ~{\rm 6s7s} \:^3{\rm S}_1$} transition. Due to
the saturation of this transition, $\sim$3 fluorescence photons
per atom exited to the \TPzero state are emitted at the probe
region. The natural width of the 649-nm transition is 70~MHz,
thus, its profile covers all transverse velocity groups ($v_x$) in
the atomic beam ($\approx8$~MHz Doppler width at 649~nm). A drift
of the laser frequency ($\sim 100$~MHz per minute) is eliminated
by locking the diode laser to a frequency-stabilized He-Ne laser
using a scanning Fabry-P\'{e}rot \'{e}talon with the scanning rate
of 25~Hz. The spectral distance between the \'{e}talon
transmission peaks from the two lasers is measured during each
scan and maintained constant within an accuracy of $\pm$3~MHz,
good enough to eliminate any degradation of the probe-region
signal.

The signals from the PMT and PD are fed into lock-in amplifiers
for frequency discrimination and averaging. A typical time of a
single spectral-profile acquisition is 20~s. The signals at the
first and the second harmonic of the electric-field modulation
frequency are registered simultaneously, which reduces the
influence of slow drifts, such as instabilities of the atomic-beam
flux. The modulation frequency is limited by several factors.
Thermal distribution of atomic velocities in the beam causes a
spread  (of $\approx$~2~ms) in the time of flight between the
interaction region and the probe region. This, along with the
finite bandwidth of the PD, leads to a reduction of the
signal-modulation contrast (see below). The choice of the
modulation frequency of 76.2~Hz is a tradeoff between this
contrast degradation and the frequent E-field reversal.

\section{Results and Analysis}
\label{Sec:Results_Analysis} In Fig.~\ref{Exp_lshape} a profile of
the B-field-split 408-nm spectral line of the $^{174}$Yb is shown.
The 649-nm-light-induced fluorescence was recorded during a single
profile scan. Statistical error bars determined directly from the
spread of data are smaller than the points in the figure. The
peculiar asymmetric line shape of the Zeeman components is a
result of the dynamic Stark effect \cite{Sta2006}.
\begin{figure}[!htb]
\resizebox{0.5\textwidth}{!}{%
  \includegraphics[bb=85 420 485 620]{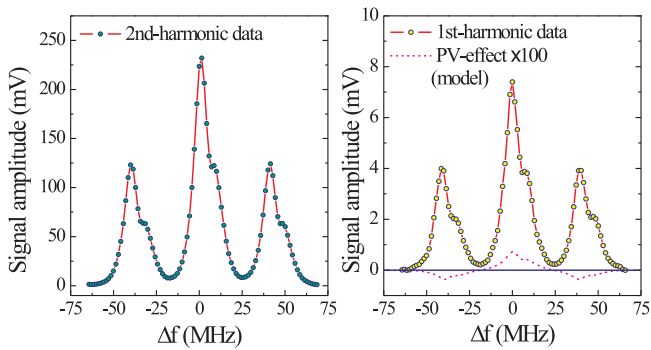}
} \caption{(color online) A profile of the B-field-split 408-nm
spectral line of $^{174}$Yb recorded at 1st- and 2nd-harmonic of
the modulation. Also shown is a simulated PV contribution in the
first-harmonic signal. $\tilde{E}$=5~kV/cm; DC offset=40~V/cm;
$\theta=\pi/4$; the effective integration time is 200~ms per
point.} \label{Exp_lshape}
\end{figure}
During a typical experimental run 100 profiles are recorded for
each combination of the magnetic field and the polarization angle
(400 profile scans in total). In order to compute the normalized
amplitude, $r_q$, of a selected Zeeman component, the actual
first-harmonic signal near the Zeeman peak is divided by the
respective second-harmonic signal and then averaged over a number
of the data points\footnote{In the normalized rate calculations
only data points having intensity higher that 1/3 of the
respective Zeeman peak are used to avoid excessive noise from
spectral regions with low signal intensity.}. Then, the
combination $\mathcal{K}$ of Eq.~(\ref{ratios}) is computed for
each profile scan followed by averaging the result over all the
scans at a given B-$\theta$ configuration. This procedure is
repeated for all four reversals, and all B-$\theta$ symmetrical
contributions, $\ratio_{1-4}$, are determined. In the present
experiment, the values of $\ratio_{2,3,4}$-terms are found to be
consistent with zero within the statistical uncertainty, which is
the same as that of the PV-asymmetry (see below).

As can be seen from Table~\ref{reversals}, terms in $\ratio_{1}$
associated with the fields imperfection are of crucial importance:
\begin{equation}
\frac{16}{\Eac}\left[e_y\left(\frac{\tilde{e}_z}{2\Eac}+\frac{\bac_x}{B}\right)+e_z\frac{\tilde{e}_y}{2\Eac}\right].\nonumber
\end{equation}
In order to measure the contribution of these terms, artificially
exaggerated E-field imperfections both static and oscillating,
$e^{ex}_z$, $e^{ex}_y$, $\tilde{e}^{ex}_y$ and $\tilde{e}^{ex}_z$,
are imposed by use of the ``correction electrodes" (see
Fig.~\ref{electrodes}), and two sets of the experiments were
performed. In the first one, a DC-voltage was applied to the
correction electrodes, and the measurements were done reversing
$e^{ex}_y$ and $e^{ex}_z$. These experiments yield values of
$\tilde{e}_y$ and $\tilde{e}_z+2\Eac \bac_x/B$. In the second set,
an AC-voltage modulated synchronously with the main E-field is
applied to the correction electrodes. In order to reverse the sign
of the parasitic terms a $\pi$-phase-shift of $\tilde{e}^{ex}_y$
and $\tilde{e}^{ex}_z$ with respect to the modulation signal is
employed by switching the wiring of the correction electrodes.
Thus, values of the DC-imperfections, $e_y$ and $e_z$, are
determined. The magnitudes of the applied electric fields and
their distributions are calculated using a 3D-numerical-model of
the interaction region. The results of the experiments are
presented in Table~\ref{stray_fields_exps}.
\begin{table}[!htb]
\caption{Results of measurements of the electric field
imperfections using artificially exaggerated AC- and
DC-components, $\tilde{e}^{ex}_{y,z}$ and $e^{ex}_{y,z}$. These
fields were generated by use of the correction electrodes,
Fig.~\ref{electrodes}. $\Eac=2000(2)$~V/cm.}
\label{stray_fields_exps}%
\begin{ruledtabular}
\begin{tabular}{cc}
\textbf{DC-Set} & \textbf{AC-Set} \\
\hline
\multicolumn{2}{c}{Exaggerated imperfections (V/cm)}\\[1ex]
$\displaystyle e^{ex}_y=-140(2)$ & $\tilde{e}^{ex}_y=-120(2)$ \\
$e^{ex}_z=20(2)$ & $\tilde{e}^{ex}_z=30(2)$ \\
\hline \multicolumn{2}{c}{Measurements (mV/cm)}\\
$\displaystyle\tilde{e}_y\frac{e^{ex}_z}{2\Eac}=16(10)$ &
$\displaystyle
e_y\frac{\tilde{e}^{ex}_z}{2\Eac}=4(5)$\\[1ex]
$\displaystyle(2\Eac\frac{\bac_x}{B}+\tilde{e}_z)\frac{e^{ex}_y}{2\Eac}=442(10)$
&
$\displaystyle e_z\frac{\tilde{e}^{ex}_y}{2\Eac}=40(5)$ \B \\[1ex]
\hline \multicolumn{2}{c}{Parasitic fields (V/cm)}\\
$\displaystyle\tilde{e}_y=3.2(2)$ & $e_y=0.5(0.6)$ \\
$\displaystyle(2\Eac\frac{\bac_x}{B}+\tilde{e}_z)=-12.6(0.3)$ &
$e_z=-1.3(0.2)$ \T
\end{tabular}%
\end{ruledtabular}
\end{table}
The net contribution of these imperfections to $\ratio_{1}$ in the
absence of the exaggerated fields is found to be \footnote{Compare
with the PV asymmetry parameter $\zeta/\beta\approx40$~mV/cm.}:
\begin{eqnarray}  \label{PV-contamination}
\displaystyle
e_y\left(\frac{\tilde{e}_z}{2\Eac}+\frac{\bac_x}{B}\right)+e_z\frac{\tilde{e}_y}{2\Eac}=&~\nonumber\\
=-2.6(1.6)_{\mbox{stat.}}(1.5)_{\mbox{syst.}}~\mbox{mV/cm.}&~
\end{eqnarray}
The systematic uncertainty comes from a sensitivity of the
numerical model of the E-field, which is used for calculating the
exaggerated fields in the interaction region, to an imperfect
approximation of the electrode-system geometry. These experiments
suggest that this field's imperfection cannot mimic the PV-effect
entirely, nevertheless, it appears to be a major source of
systematic uncertainty impacting the accuracy of the PV-asymmetry
measurements. The most prominent contribution is given by a
combined effect of the parasitic components of the electric field
and the non-zero projection of the leading magnetic field on the
direction of the electric field:
$e_y(\tilde{e}_z/2\Eac+\bac_x/B)$. The PV-asymmetry parameter,
$\zeta/\beta$ is obtained from the measured value of $\ratio_{1}$
by compensating for the influence of these magnetic- and
electric-field imperfections, Eq.~(\ref{PV-contamination}).

There is another effect that cannot, by itself, mimic the
PV-asymmetry, but needs to be taken into account for proper
calibration. This effect is related to the E-field modulation
implemented in the present experiment. The atoms are excited to
the metastable state, \TPZero, by the light beam in the
interaction region and then travel $\sim$20~cm until they are
detected downstream in the probe region. Due to the spread in the
time-of-flight between the interaction and probe regions, the
phase mixing leads to a reduction of the signal modulation
contrast at the probe region, and it depends on the
signal-modulation frequency. Since the signal comprises two
time-scales of interest, first- and second-harmonic of the E-field
modulation, the contrast reduction is different for the two.
Therefore, the ratio of the signal modulation amplitudes, $r_M$,
on which we base the PV-asymmetry observation, appear altered in
the probe region compared to what it would be at the interaction
region. The amplitude combination, $\mathcal{K}$, and, therefore,
the PV-parameter, $\zeta/\beta$, are similarly affected. In our
data analysis, a correction coefficient, $C_0$, is introduced,
which has been calculated theoretically:
\begin{eqnarray}  \label{phase-mixing}
\left[\frac{\zeta}{\beta}\right]_{\mbox{probe~reg.}}&=&C_0\left[\frac{\zeta}{\beta}\right]_{\mbox{real}}.\nonumber
\end{eqnarray}
Under present experimental conditions, this coefficient, $C_0$, is
found to be 1.028(3), and the measured PV parameter is corrected
accordingly. Principles of its derivation are given in
Appendix~\ref{Ap:phase_mixing}.
\begin{figure}[!htb]
\resizebox{0.5\textwidth}{!}{%
  \includegraphics[bb=260 270 520 410]{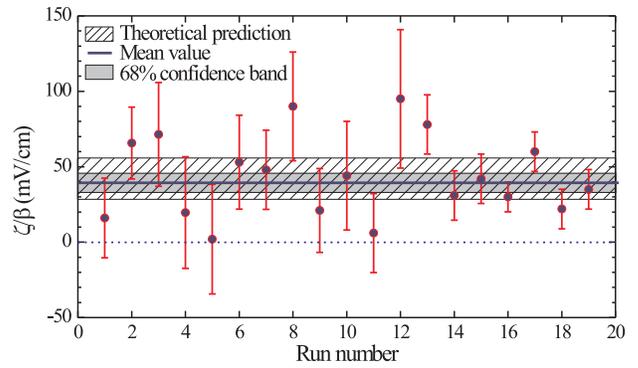}
} \caption{(color online) The PV interference parameter $\zeta
/\beta$. Mean value:
$39(4)_{\text{stat.}}(3)_{\text{syst.}}~\text{mV/cm}$, $|\zeta
|=8.7\pm 1.4\times 10^{-10}$~ea$_0$.} \label{APV}
\end{figure}

In Fig.~\ref{APV}, the PV interference parameter $\zeta /\beta$ is
shown as determined in 19 separate runs ($\sim$60 hours of
integration in total). Its mean value is
$39(4)_{\text{stat.}}(3)_{\text{syst.}}~\text{mV/cm}$, which is in
agreement with the theoretical predictions \cite{Por95,Das97}. The
value of the PV parameter was extracted using the expression given
in the first column of Table~\ref{reversals}, taking into account
the calibration correction $C_0$. Thus, $|\zeta |=8.7\pm 1.4\times
10^{-10}$~ea$_0$, which is the largest APV amplitude observed so
far (here we used $|\beta |= 2.24_{-0.12}^{+0.07}\times
10^{-8}~\text{ea}_0$/(V/cm) \cite{Bow99,Sta2006}).

The sign of the PV interference parameter $\zeta/\beta$ is found
by comparing the measurements with the theoretical model of the
transition rates employing the field geometry shown in
Fig.~\ref{apparatus}. The direction and, thus, the signs of the
electric and magnetic fields as well as the polarization angle
$\theta$ were calibrated prior to the PV measurements. Special
care was taken of detecting parasitic phase shifts in the lock-in
amplifier. A signal from an arbitrary function direct digital
synthesis (DDS) generator simulating the output of the probe
region photodetector was fed into the amplifier. The signal is
comprised of a sum of two sinusoidal waveforms, one frequency
doubled, attenuated, and phase shifted with respect to the other.
Results of the signal parameters measurement from the lock-in,
such as the first-to-second harmonic amplitude ratio, relative
phase shift and its sign, are compared to those used in the DDS
generator to simulate the signal. The difference in the measured
and generated amplitude ratio is found to be below 0.01\%, and the
relative phase shift is detected within $\pm 1.5^{\circ}$. No
relative sign flips between the first- and second-harmonic
amplitudes were detected.

\section{Error budget}
\label{Sec:Error_budget} The present measurement accuracy is not
yet sufficient to observe the isotopic and hyperfine differences
in the PV amplitude, which requires an accuracy better than
$\approx 1\%$ for PV amplitude in a single transition
\cite{nSkin,AnapoleKozlov,Das99}. In the present apparatus the
signal levels achieved values high enough to reach the
signal-to-noise ratio (SNR) of $2/\sqrt{\tau(\mbox{s})}$ for the
PV asymmetry if the noise were dominated by the
fluorescence-photon shot-noise ($\tau$ is the integration time).
This is good enough to reach the sub-percent accuracy in a few
hours of integration. However, a number of additional factors
limit the accuracy.

One of the most important noise sources is the fluctuations of the
modulating- and DC-field parameters during the experiment. The
first- and the second-harmonic signals depend differently on the
modulating electric field amplitude, \Eac, and the DC-bias, thus,
a noise in the electronics controlling the fields contaminates the
first-to-second harmonic ratio directly. A substantial effort was
made to cope with this problem: a home-built high-voltage
amplifier used in the first 13 runs was replaced by a commercial
Trek 609B amplifier and a circuit controlling the DC-bias was
upgraded. This allowed us to control the DC-bias and \Eac with
mV-scale accuracy that would make the SNR to approach the
shot-noise limit if this were the only source of the noise. As
seen in Fig.~\ref{APV}, the last six measurements exhibit higher
accuracy than the rest. These are the runs after the HV-system
upgrade. However, the present SNR of $\approx
0.03/\sqrt{\tau(\mbox{s})}$ is worse by almost two orders of
magnitude than the shot-noise limit.

There are other fluctuations in the system parameters, such as
light intensity fluctuations in the PBC, fluctuations of the
spectral position of the PBC resonance with respect to the
frequency reference, and noise in the detection system. All of
them contribute to the noise in the first- and the second-
harmonic signals but we found that such noise largely canceled in
the ratio $r_M$.

However, there is a noise source, which is not canceled in the
ratio. The following experiments demonstrated that this noise
source is related to frequency excursions of the Fabry-P\'{e}rot
\'{e}talon serving as the frequency reference for the optical
system. In these experiments the excitation light was frequency
tuned to a wing of the atomic resonance, and the first- and
second-harmonic signals were recorded without scanning over the
resonance. Then, the same was done when the spectral position was
set at the peak of the resonance, and a change of the SNR for the
harmonics ratio was determined. These experiments were performed
using the upgraded HV-system. Results of the measurements are
presented in Fig.~\ref{FP-noise}.
\begin{figure}[!htb]
\resizebox{0.5\textwidth}{!}{%
  \includegraphics[bb=170 355 445 540]{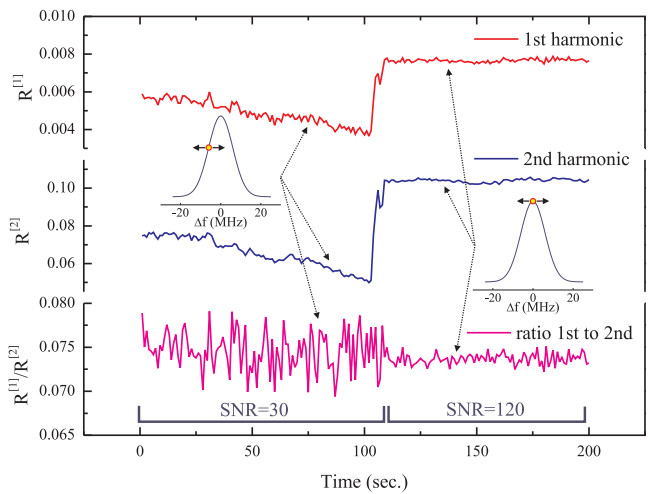}}
  \caption{(color online) Impact of the frequency excursions
  of the Fabry-P\'{e}rot \'{e}talon on the noise level in the harmonics ratio.
  A change in the noise level when the optical system was
  tuned from the wing of the atomic resonance to its peak is shown.
  In the inserts above the excitation light spectral position is
  shown schematically with respect to the atomic resonance.
  Arrows denote the fluctuations.
  } \label{FP-noise}
\end{figure}
For a shot-noise-limied signal, the SNR at the peak of the
resonance is expected to be a factor of about $\sqrt{2}$ higher
than at the wing due to larger signal. It was found, however, that
the SNR went up by a factor of 4 by tuning from the wing to the
peak of the resonance. This demonstrates that the main source of
noise is not photon statistics but fluctuations in the spectral
reference. Indeed, the frequency excursions at the wing of the
spectral line produces substantially more intensity noise due to a
steeper spectral slope than that at the peak, where the slope is
nominally zero. It must be emphasized, that in the case of slow
frequency excursions (compared to the E-field modulation period),
the noise in the first- and the second-harmonic channels would be
canceled in the ratio. However, fast excursions can generate noise
in the signal ratio.

The factors affecting the measurement accuracy mentioned above
have an impact on the statistical error of the result. The present
systematic errors (summarized in Table~\ref{error_budget}) has
nearly the same significance as the statistical one and also
comprises a number of factors.

One of the most significant factors is the uncertainty in the
field-imperfection contributions, Eq.~(\ref{PV-contamination}).
This uncertainty is mostly due to statistical factors such as
laser drifts, nevertheless, it provides an offset to the
PV-parameter. Since the measurement of this contribution is
actually the same measurement as the PV-effect, any improvements
of the stability reduces the overall systematic uncertainty. We
would like to emphasize also that Eq.~(\ref{PV-contamination})
represents a mean value of the imperfection contribution over
numerous experiments averaging over possible fluctuations of the
field-imperfection contribution. These fluctuations may be
partially responsible for the variance in the PV-parameter, and,
thus, the statistical uncertainty of its value. This fact
demonstrates that the elimination of the field-imperfections is an
essential requirement for improving the overall accuracy of the
experiments.

Another significant source of the systematic uncertainty is the
uncertainty in the value of the electric field in the interaction
region. While the voltage applied to the E-field plates and the
correction electrodes is controlled precisely, the actual E-field
value used in the PV parameter determination depends on the
accuracy of the numerical modeling of the electric-field
distribution in the particular geometry. There are two factors in
the model contributing to the uncertainty: finite accuracy of
measurements of the interaction region geometrical parameters, and
the imperfect approximation of the geometry in the numerical
simulation.

However, while this systematic uncertainty plays a significant
role for measurements of the PV parameter of a single isotope, for
the isotope ratios this uncertainty will cancel (or will be
substantially reduced), if the measurements observing different
isotopes are performed without changing the E-field geometry. The
same is true for the calibration parameter, $C_0$, which also
cancels in the isotope ratios.

There are other, rather minor, factors contributing to the
systematic uncertainty, for example, a finite accuracy of the
polarization angle flip, errors in the lock-in amplifiers, a
finite dynamic range of the lock-ins etc. The net contribution of
these factors is found to be $\lesssim 1\%$.

A summary of the systematic error budget is presented in
Table~\ref{error_budget}.
\begin{table}[!htb]
\caption{List of factors contributing to the systematic
uncertainty of the PV parameter, $\zeta/\beta$.}
\label{error_budget}%
\begin{ruledtabular}
\begin{tabular}{l|l}
\textbf{Factor} & \textbf{Uncertainty (\%)}\\
\hline%
$\tilde{E}$ value: &~\\
~~~geometry & 5\\
~~~numerical modeling & 3\\
E-field imperfections & 5\\
Phase mixing & 0.5\\
Other & 1 \\
\hline%
\textbf{Total} (in quadrature) & \textbf{8}
\end{tabular}%
\end{ruledtabular}
\end{table}

\section{Towards measurement of the isotope ratios}
\label{Sec:Future} As pointed out above, a goal of the future
measurements of the parity-violation effects in ytterbium is
observing a difference in the PV effect between different
isotopes. The net uncertainty of the PV parameter of a single
isotope must be better than 1\% based on the theoretical
predictions. To this end, a program of the apparatus upgrades and
improvements is developed. Besides general improvements of the
stability of the system parameters, increase of the signal levels,
suppression of the electronics noise etc., the main focus is on
elimination of the frequency excursions of the frequency
reference, which is a major source of the statistical noise.
Improving the statistical uncertainty will contribute to more
precise measurement and control of the E-field-imperfection
contribution to the systematic part of the uncertainty. The latter
is another high-priority improvement essential for reaching the
goal.

In the future apparatus, the referencing of the optical system to
the Fabry-P\'{e}rot \'{e}talon will be replaced by locking the
system to a femtosecond frequency comb that will be available
shortly. The impact of the E-field imperfection is planned to be
substantially suppressed by redesigning of the interaction region
to provide more uniform and controlled electric field
distribution. Until now, no scientific or technical obstacles were
discovered preventing us to improve the apparatus to the desired
level of sensitivity.

\section{Acknowledgements}
\label{Acknowledgements} The authors acknowledge helpful
discussions with and important contributions of M. A. Bouchiat, C.
J. Bowers, E. D. Commins, B. P. Das, D. DeMille, A. Dilip, S. J.
Freedman, J. S. Guzman, G. Gwinner, M. G. Kozlov, S. M. Rochester,
and M. Zolotorev. This work has been supported by NSF.

\appendix
\section{Derivation of transition amplitudes}
\label{Ap:amplitudes}%
The total Hamiltonian (before including light-atom interactions
and assuming $\Bfield$ is along $\zhat$ ) can be written as
\begin{equation}\label{eq:Htot}
H = \HAtomic + \HZeeman + \HStark + \HAPV,
\end{equation}
where $\HAtomic$ is the atomic Hamiltonian and $\HZeeman$,
$\HStark$, and $\HAPV$ represent the contributions from the static
magnetic field $\Bfield$, the static electric field $\Efield$, and
the parity non-conserving weak interaction, respectively.  Here
\begin{equation}\label{eq:HZeeman}
\HZeeman = -\Mdipole\cdot\Bfield = g\muB\J\cdot\Bfield = g\muB
J_z\bfield,
\end{equation}
where $\Mdipole=-g\muB\J$ is the magnetic dipole moment of the
atom, $g$ is the Land\'e factor, $\muB$ is the Bohr magneton, and
$\J$ is the angular-momentum operator. Similarly,
\begin{equation}\label{eq:HStark}
 \HStark = -\Edipole\cdot\Efield =
 -\edipole_i\efield_i,
\end{equation}
where $\Edipole$ is the atomic electric-dipole operator. Finally,
\begin{equation}\label{eq:HPNC}
\HAPV = i\kq{H}{0}{0},
\end{equation}
where $\kq{H}{0}{0}$ is a scalar operator.  Summation over
repeated indices is assumed.

In the presence of a strong magnetic field, that is, when Zeeman
splitting dominates Stark-shifts, it is useful to think of $H_1
\equiv \HStark + \HAPV$ as a perturbation to $H_0 \equiv
\HAtomic+\HZeeman$. In this case, the $LS$-coupled states
$\ket{\SLJ;M}$, such as $\ket{\TDone;M}$ and $\ket{\SSzero;0}$,
are eigenstates of the unperturbed Hamiltonian $H_0$. Then the
first-order perturbation theory can be used to determine the
eigenstates of the total Hamiltonian:
\begin{equation}\label{eq:threeDoneBar}
\ket{\overline{a}} = \ket{a} + \sum_{a'}\frac{
\ket{a'}\bra{a'}H_1\ket{a}}{\energy(a)-
\energy(a')},
\end{equation}
where $\omega(a)$ is the energy of state $\ket{a}$.
(Perturbed eigenstates are denoted using an overbar.)

The electric-dipole amplitude for the optical transition of
interest is
\begin{equation}
\AEONE_M =
\bra{\overline{\TDone;M}}(-\Edipole\cdot\Elight)\ket{\overline{\SSzero}}
\equiv \ASTARK_M + \APV_M,
\end{equation}
where
\begin{align}
\label{eq:AStark_1}%
\ASTARK_M = & \sum_{a'}
\frac{\bra{\TDone;M}\Edipole\cdot\Efield\ket{a'}
\bra{a'}\Edipole\cdot\Elight\ket{\SSzero}}{\energy(\TDone)-\energy(a')}\nonumber\\
&+\sum_{a'}\frac{\bra{\TDone;M}\Edipole\cdot\Elight\ket{a'}
\bra{a'}\Edipole\cdot\Efield\ket{\SSzero}}{\energy(\SSzero)-\energy(a')},
\end{align}
and
\begin{align}
\APV_M = &
\sum_{a'}\frac{\bra{\TDone;M}i\kq{H}{0}{0}\ket{a'}
\bra{a'}\Edipole\cdot\Elight\ket{\SSzero}}{\energy(\TDone)-\energy(a')}\nonumber\\
&-\sum_{a'}
\frac{\bra{\TDone;M}\Edipole\cdot\Elight\ket{a'}
\bra{a'}i\kq{H}{0}{0}\ket{\SSzero}}{\energy(\SSzero)-\energy(a')}.
\end{align}

The Stark amplitude can be written as
\begin{equation}
\ASTARK_M = T_{ij}\bra{\TDone;M}U_{ij}\ket{\SSzero},
\end{equation}
where $T_{ij} = \efield_i\elight_j$ and
\begin{align}
U_{ij} = \sum_{a'}
\frac{\edipole_i\ket{a'}\bra{a'}\edipole_j}{\energy(\TDone)-\energy(a')}
+
\frac{\edipole_j\ket{a'}\bra{a'}\edipole_i}{\energy(\SSzero)-\energy(a')}.
\end{align}
Let $\kq{T}{k}{q}$ and $\kq{U}{k}{q}$ represent the irreducible
spherical components of the tensors $T_{ij}$ and $U_{ij}$.  Then
$T_{ij}U_{ij} = (-1)^q\kq{T}{k}{-q}\kq{U}{k}{q}$ and Eq.
(\ref{eq:AStark_1}) becomes
\begin{align}\label{eq:ASTARK}
\ASTARK_M &=
(-1)^q\kq{T}{k}{-q}\bra{\TDone;M}\kq{U}{k}{q}\ket{\SSzero}\nonumber\\&
= (-1)^q\kq{T}{k}{-q} \frac{\reducedME{\TDone}{U}{k}{\SSzero}}
{\sqrt{3}}\clebsch{0}{0}{k}{q}{1}{M}\nonumber\\
&=
i\beta(-1)^q(\Efield\times\Elight)_{-q}\clebsch{0}{0}{1}{q}{1}{M}.
\end{align}
Here $\beta$ is the vector Stark transition polarizability and
defined by
\begin{equation}
\beta \equiv \frac{1}{\sqrt{6}}\reducedME{\SSzero}{U}{1}{\TDone}.
\end{equation}
To derive Eq. (\ref{eq:ASTARK}), we used
$\clebsch{0}{0}{k}{q}{1}{M} = \delta_{k1}\delta_{qM}$ and
$\kq{T}{1}{-q}
=\sum_{q_1,q_2}\clebsch{1}{q_1}{1}{q_2}{1}{-q}E_{q_1}\elight_{q_2}
= (i/\sqrt{2})(\Efield\times\Elight)_{-q}$.

For the parity-violating contribution to the E1 transition
amplitude we can likwise write
\begin{equation}
\APV_M = i\elight_i\bra{\TDone;M}W_i\ket{\SSzero},
\end{equation}
where
\begin{equation}
W_i = \sum_{a'}
\frac{\kq{H}{0}{0}\ket{a'}\bra{a'}\edipole_i}{\energy(\TDone)-\energy(a')}
- \sum_{a'}
\frac{\edipole_i\ket{a'}\bra{a'}\kq{H}{0}{0}}{\energy(\SSzero)-\energy(a')}.
\end{equation}
Let $\kq{\elight}{1}{q}$ and $\kq{W}{1}{q}$ represent the
spherical components of the vectors $\elight_i$ and $W_i$,
respectively.  Then
$\elight_iW_i=(-1)^q\kq{\elight}{1}{-q}\kq{W}{1}{q}$ and we have
\begin{align}
\APV_M &= i(-1)^q\kq{\elight}{1}{-q}\bra{\TDone;M}\kq{W}{1}{q}\ket{\SSzero}\nonumber\\
&=i(-1)^q\kq{\elight}{1}{-q}
\frac{\reducedME{\TDone}{W}{1}{\SSzero}}
{\sqrt{3}}\clebsch{0}{0}{1}{q}{1}{M}\nonumber\\
&= i\zeta(-1)^q\kq{\elight}{1}{-q}\clebsch{0}{0}{1}{q}{1}{M}.
\end{align}
Here $\zeta$ is given by
\begin{equation}
\zeta \equiv \frac{1}{\sqrt{3}}\reducedME{\TDone}{W}{1}{\SSzero}.
\end{equation}

\section{Characterization of the PBC mirrors}
\label{Ap:PBC_mirrors} The finesse of the cavity is measured using
the cavity-ring-down method \cite{ring_down}. The laser beam is
sent through a Pockels cell (Cleveland Crystals Inc. QX 1020
Q-Switch) and a polarizer before entering the cavity. The
polarizer is aligned with the laser polarization so that the light
is transmitted when there is no voltage applied to the Pockels
cell. A high-voltage pulse generator is used to send a fast step
signal (30-ns wavefront) to the Pockels cell which rotates the
polarization of the light so that it is not transmitted through
the polarizer. The laser frequency is locked to the resonance
frequency of the cavity, and then the Pockels cell is switched
into the non-transmitting state, causing a fast interruption of
the laser power. The subsequent decay of the light inside the
cavity is monitored with a fast photodiode (50-MHz bandwidth)
measuring the power transmitted through the back mirror of the
cavity. The signal is fit to an exponential decay. The decay time
is related to the finesse of the cavity ($\mathcal{F}$) by
\begin{equation}
\mathcal{F}=\frac{\pi c}{L}\tau, \nonumber
\end{equation}
where $c$ is the speed of light, $L$ is the cavity length, and
$\tau$ is the intensity decay time. An example of the PBC
transmission signal and its fit are shown in
Fig.~\ref{Fig:ringdown}.
\begin{figure}[!htb]
\resizebox{0.5\textwidth}{!}{
  \includegraphics{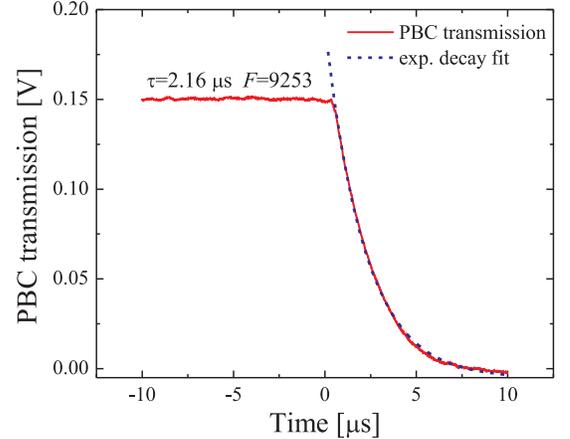}}
  \caption{(color online) Application of the cavity-ring-down
  method for the determination of the finesse of PBC.
  } \label{Fig:ringdown}
\end{figure}

Following the analysis discussed in \cite{PBCmirrors}, if we
denote the transmission of mirrors 1 and 2 by $T_1$ and $T_2$,
respectively, and the absorption+scatter loss per mirror as
$l_{1,2}=(A+S)_{1,2}$, then the total cavity losses
$\mathcal{L}=T_1+T_2+l_1+l_2$ determine the finesse $\mathcal{F}$:
\begin{equation}
\label{finesse}%
\mathcal{F}=\frac{2\pi}{T_1+T_2+l_1+l_2}.
\end{equation}
Information on the transmission of the mirrors discriminated from
the A+S losses can be obtained using the measured value of the
finesse and the power transmitted trough PBC, $P_{tr}$:
\begin{equation}
\label{transmission}%
\frac{P_{tr}}{\epsilon P_{in}}=4 T_1 T_2
\left(\frac{\mathcal{F}}{2\pi}\right)^2,
\end{equation}
where $P_{in}$ is the input power, and $\epsilon$ is a
mode-matching factor. For two arbitrary mirrors, for which neither
$T_{1,2}$ nor $l_{1,2}$ are known the Eq.
(\ref{finesse},\ref{transmission}) do not provide a solution,
since a number of variables exceeds the number of equations.
Nevertheless, for two mirrors from the same coating run when one
can assume that $T_1=T_2=T$ and $l_1=l_2=l$, the equations
(\ref{finesse},\ref{transmission}) become
\begin{align}
\label{PBC_eq_system}%
\mathcal{F}&=\frac{\pi}{T+l}. \nonumber \\
\frac{P_{tr}}{\epsilon P_{in}}&=4 T^2
\left(\frac{\mathcal{F}}{2\pi}\right)^2, \nonumber
\end{align}
and for known mode-matching factor $\epsilon$ the parameters of
the mirrors ($T$ and $l$) can be determined. The factor $\epsilon$
depends on the geometry of the cavity, and is assumed to stay
constant upon replacing of the mirrors, if the geometry of the
input laser beam and the configuration of the PBC are unchanged.
This gives the possibility to calibrate this factor by using a
mirror set for which the transmission is known. We used for this
purpose the mirror set purchased from Advanced Thin Films, Inc.,
for which reliable data on the transmission of the mirrors is
provided by the supplier. By measuring the finesse of the PBC
comprised of these mirrors and the ratio of the
transmitted-to-input power, the mode-matching factor and the A+S
mirror losses $l$ are found. This set is not an actual mirror set
that was used in the PV experiment, nevertheless, the parameters
of other mirrors were determined by replacing one mirror in the
``reference" set by the ``test" mirror, parameters of which are
sought. The geometry of the cavity was unchanged during the
replacement. This tactic allows for the measurement of parameters
of any arbitrary mirror.

\section{Impact of the phase mixing effect on the harmonics ratio}
\label{Ap:phase_mixing} Atoms undergo the \SSZeroToTDOne
transition in the interaction region where they are illuminated by
408-nm light and are exposed to the static magnetic field and the
oscillating electric field $E(t)$. Excited atoms then
spontaneously decay from the \TDOne state to the metastable
\TPZero state.  The population of \TPZero is proportional to the
transition rate $R_M$ for $M=0,\pm1$.  Without loss of generality,
we assume that the constant of proportionality is equal to one.

The rate $R_M$ is measured in the probe region.  The probe region
is located a distance $d \approx $ 20 cm away from the interaction
region. Therefore, an atom that arrives at the detection region at
time $t$ experienced an electric field with magnitude $E(t-d/v_z)$
in the interaction region, where $v_z$ is the atom's speed and
$d/v_z$ is the amount of time required for the atom to travel a
distance $d$.

Because some atoms travel faster or slower than others, the
detection region is full of atoms that have each experienced a
different electric field while in the interaction region.  Each
atom contributes to the total rate and hence the observed rate
$\overline{R}_M$ is the thermal average of every contribution:
\begin{eqnarray}\label{eq:Rbar1}
\overline{R}_M(t;\omega,d,\vzmp) =
\int_0^{\infty}R_M(t-d/v_z)f(v_z;\vzmp)\,dv_z,
\end{eqnarray}
where
\begin{eqnarray}
f(v_z;\vzmp)\,dv_z = 2(v_z/\vzmp)^3e^{-(v_z/\vzmp)^2}dv_z/\vzmp,
\end{eqnarray}
is the probability for an atom to have speed between $v_z$ and
$v_z+dv_z$.  Here $\vzmp = \sqrt{2\kB T/m}= 2.9\times10^4$ cm/s is
the thermal speed, $T\approx$ 873 K is the oven temperature, and
$m = 161$ $\mathrm{GeV}/c^2$ is the atomic mass of Yb.

It is convenient to introduce the dimensionless variables
$x=v_z/\vzmp$ and $\tau=\omega t$, and the dimensionless parameter
$\alpha = \omega d/\vzmp$. Then the average rate
$\overline{R}_M(t;\omega,d,\vzmp)\rightarrow\overline{R}_M(\tau;\alpha)$
depends only on the dimensionless quantities $\alpha$ and $\tau$,
and Eq. (\ref{eq:Rbar1}) becomes
\begin{align}
\overline{R}_M(\tau; \alpha) =\;& \ModRate{0}{M} +
\ModRate{1}{M}|I(\alpha)|\cos(\tau+\Arg[I(\alpha)]) \nonumber\\
&+\ModRate{2}{M}|I(2\alpha)|\cos(2\tau+\Arg[I(2\alpha)]),
\end{align}
with
\begin{eqnarray}
I(\alpha) \equiv \int_0^{\infty}e^{-i\alpha/x}f(x;1)\,dx.
\end{eqnarray}
Note that $|I(\alpha)|\rightarrow0$ as $\alpha\rightarrow\infty$
whereas $|I(\alpha)|\approx 1$ when $\alpha < 1$.  This places a
limit on the modulation frequency:  We require that $\omega <
\vzmp/d = 2\pi\times$ 230 Hz in order to avoid a significant
decrease in signal.

The lock-in amplifier receives an input signal proportional to
$\overline{R}_M$ and returns two output signals $\kq{S}{[1]}{M}$
and $\kq{S}{[2]}{M}$ corresponding to the first and second
harmonic components, respectively.  This process can be modeled as
\begin{align}
\kq{S}{[n]}{M}(\phi_n;\alpha) &=
\frac{1}{\pi}\int_0^{2\pi}\overline{R}_M(\tau;\alpha)\cos(n\tau+\phi_n)\,d\tau
\nonumber\\ &= \ModRate{n}{M}
|I(n\alpha)|\cos(\Arg[I(n\alpha)]+\phi_n),
\end{align}
where the phases $\phi_{1,2}$  of the lock-in amplifier are chosen
to maximize the signals $S^{[1,2]}_{M}$.  That is,
\begin{eqnarray}
\phi_n = \phi_n(\alpha)\equiv -\Arg[I(n\alpha)].
\end{eqnarray}
Our measurement $s_M$ is the ratio of the first- and second-
harmonic signals:
\begin{eqnarray}
s_M =
\frac{\kq{S}{[1]}{M}(\phi_1;\alpha)}{\kq{S}{[2]}{M}(\phi_2;\alpha)}
= \frac{\ModRate{1}{M}|I(\alpha)|}{\ModRate{2}{M}|I(2\alpha)|}=
r_M\times C(\alpha),
\end{eqnarray}
where $C(\alpha)\equiv|I(\alpha)|/|I(2\alpha)|$ is the
\emph{correction factor}.  Therefore, we must further divide the
ratio $s_M$ of observed output signals by $C(\alpha)$ to measure
the ratio $r_M$.

The correction factor $C(\alpha)$ and the optimal lock-in phases
$\phi_{1,2}(\alpha)$ inherit dependence on the modulation
frequency ($\omega= 2\pi\times76.2$ Hz), the distance between
interaction and detection regions ($d \approx 20$  cm), and the
oven temperature ($T\approx$ 873 K) through the parameter
$\alpha$:
\begin{eqnarray}
\alpha = \frac{\omega\, d}{\sqrt{2\kB T/m}} = 0.33(2),
\end{eqnarray}
where the uncertainty in $\alpha$ is given by
\begin{eqnarray}
\delta\alpha = \alpha\sqrt{(\delta T/2T)^2+(\delta d/d)^2},
\end{eqnarray}
for $\delta T\approx$ 50 K and $\delta d\approx$ 1 cm. The
correction factor can be computed numerically and has a value
\begin{eqnarray}
C_0 = C(\alpha) = 1.028(3),
\end{eqnarray}
with uncertainty given by $\delta C_0 =
|C'(\alpha)|\,\delta\alpha$.  Likewise, the lock-in phases have
the following values
\begin{eqnarray}
\phi_{10} = \phi_1(\alpha) = 16(1)^{\circ},\quad \phi_{20} =
\phi_2(\alpha) = 33(2)^{\circ},
\end{eqnarray}
where  $\delta\phi_{n0} = |\phi_n'(\alpha)|\,\delta\alpha$.

In order to understand the impact of imperfect phase selections,
we include the effects of slight deviations from the optimal phase
$\phi_n(\alpha)$ by taking
\begin{eqnarray}
\phi_n\rightarrow\phi_n(\alpha)+\varphi_n,
\end{eqnarray}
where $\varphi_n\approx0$ represents a small deviation.  Then the
correction factor becomes
\begin{eqnarray}
C(\alpha) \rightarrow \tilde{C}(\alpha,\varphi_1,\varphi_2) =
C(\alpha)\times \frac{\cos(\varphi_1)}{\cos(\varphi_2)},
\end{eqnarray}
and hence $\tilde{C}_0 = \tilde{C}(\alpha,0,0) = C(\alpha) =
C_0$.  The uncertainty in the correction factor becomes
\begin{eqnarray}
\delta C_0 \rightarrow \delta \tilde{C}_0 = \sqrt{\delta C_0^2 +
\delta\varphi_1^4 + \delta\varphi_2^4},
\end{eqnarray}
where $\delta\varphi_n$ is the uncertainty in the deviation
$\varphi_n$.  To derive this expression, we estimated the partial
uncertainty in $\tilde{C}_0$ due to $\varphi_n$ by
$\partial^2_{\varphi_n}\tilde{C}(\alpha,\varphi_1,\varphi_2)\,\delta\varphi_n^2$.

To estimate the uncertainty $\delta\varphi_n$, we assume that we
are within about $1^{\circ}$ of the optimal phase.  This choice is
consistent with the magnitude of the uncertainty in the optimal
phases $\phi_{10}$ and $\phi_{20}$.  Therefore, we will take
$\delta\varphi_n = \delta\phi_{n0}$ to be the accuracy with which
we can select the lock-in phases.  Then $\delta\varphi_1 = 0.02$,
$\delta\varphi_2 = 0.03$, and
\begin{eqnarray}
\delta \tilde{C}_0 = 0.0031 \approx 0.0029 = \delta C_0.
\end{eqnarray}
Hence small deviations (on the order of $1^{\circ}$) have a
negligible effect on the uncertainty in the correction factor.

\bibliography{Yb_APV_systematics}

\end{document}